\documentstyle[12pt]{article}                                      
                                      
\newlength{\dinwidth}                                      
\newlength{\dinmargin}                                      
\def\lapproxeq{\lower .7ex\hbox{$\;\stackrel{\textstyle <}{\sim}\;$}}    
\def\gapproxeq{\lower .7ex\hbox{$\;\stackrel{\textstyle >}{\sim}\;$}}     
\def\be{\begin{equation}}                                      
\def\ee{\end{equation}}                                      
\def\bea{\begin{eqnarray}}                                      
\def\eea{\end{eqnarray}}                                     
\def\eV{\rm eV}

\begin{document}                                      
\titlepage                                      
\begin{flushright}                                      
DTP/98/78 \\                                      
UCRHEP-T240\\
November 1998 \\                                      
\end{flushright}                                      
                                      
\vspace*{2cm}                                      
                                      
\begin{center}                                      
{\Large \bf Minimal see-saw model for atmospheric and} \\             
                        
\vspace*{0.5cm}                        
{\Large \bf solar neutrino oscillations}                                      
                                      
\vspace*{1cm}                                      
Ernest Ma$^{a}$ and D.P.~Roy$^{b,c}$\\                                      
                                     
\vspace*{0.5cm}                                      
$^a$ Department of Physics, University of California, Riverside, 
California 92521, USA. \\                       
$^b$ Department of Physics, University of Durham, Durham, DH1 3LE, UK. \\   
$^c$ Tata Institute of Fundamental Research, Homi Bhabha Road, Bombay 400005, 
India.                                     
\end{center}                                      
                                      
\vspace*{2cm}                                      
                                      
\begin{abstract}                                      
We present a minimal see-saw model based on an extension of the standard model 
(SM) which includes an additional U(1), with gauge charge $B - \frac{3}{2} 
(L_\mu + L_\tau)$.  Requirement of anomaly cancellation implies the 
existence of two right-handed singlet neutrinos, carrying this gauge charge, 
which have normal Dirac couplings to $\nu_\mu$ and $\nu_\tau$ but suppressed 
ones to $\nu_e$.  Assuming the U(1) symmetry breaking scale to be 
10$^{12-16}$~GeV, this model can naturally account for the large (small) 
mixing solutions to the atmospheric (solar) neutrino oscillations. 
\end{abstract}                                     
                             
\newpage                        
\baselineskip 24pt             
Super-Kamiokande data have recently provided convincing evidence for 
atmospheric neutrino oscillations \cite{SKC1} as well as confirmed earlier 
results on solar neutrino oscillations \cite{SKC2}.  The atmospheric 
neutrino oscillation data seem to require a large mixing angle between 
$\nu_\mu$ and $\nu_\tau$, 
\be                      
\sin^2 2 \theta_{\mu \tau} \; > \; 0.82                      
\label{eq:a1}                      
\ee                      
and                      
\be                      
\Delta M^2 \; = \; (0.5 - 6) \: \times \: 10^{-3} \eV^2.                      
\label{eq:a2}                      
\ee                      
On the other hand, the solar neutrino oscillation data can be explained by 
the small mixing angle matter enhanced (MSW) \cite{LW} solution between 
$\nu_e$ and a combination of $\nu_\mu/\nu_\tau$ with \cite{BKS} 
\be                      
\sin^2 2 \theta_{e - \mu/\tau} \; = \; 10^{-2} \: - \: 10^{-3}                      
\label{eq:a3}                      
\ee                      
and                      
\be                      
\Delta m^2 \; = \; (0.5 - 1) \: \times \: 10^{-5} \eV^{2}.                      
\label{eq:a4}                      
\ee                      
This represents the most conservative solution to the solar neutrino anomaly 
although one can get reasonably good solutions with large mixing angle MSW 
and vacuum oscillations as well.  One would naturally expect a near-maximal 
mixing between $\nu_\mu$ and $\nu_\tau$ (\ref{eq:a1}), as required by the 
atmospheric neutrino data, if they were almost degenerate Dirac partners 
with a small mass difference given by (\ref{eq:a2}).  In the context of a 
three neutrino model however, the solar neutrino solution (\ref{eq:a4}) 
would then require the $\nu_e$ to show a much higher level of degeneracy 
with one of these states, which is totally unexpected.  Therefore, it is 
more natural to consider the three neutrino mass eigenstates as non-degenerate 
with 
\be                     
m_1 \; = \; (\Delta M^2)^{1/2} \; \simeq \; 0.05~\eV, \quad\quad m_2 \; = \; (\Delta                     
m^2)^{1/2} \; \simeq \; 0.003~\eV, \quad\quad m_3 \ll m_2.                    
\label{eq:a5}                     
\ee                    
There is broad agreement on this point in the current literature on 
neutrino physics \cite{BA}--\cite{SFK}, much of which is focused on the 
question of reconciling this hierarchical structure of neutrino masses 
with at least one large mixing angle (\ref{eq:a1}).  It may be noted here 
that in a minimal scenario one needs only two neutrino masses with $m_3 
\rightarrow 0$, since it has no relevance for atmospheric or solar neutrino 
oscillations.
                    
The cannonical mechanism for generating neutrino masses is the so-called 
see-saw model, containing heavy right-handed singlet neutrinos \cite{GMRS}, 
which induce small hierarchical masses for the doublet neutrinos.  The 
standard see-saw model is based on a U(1) extension of the standard-model 
(SM) gauge group SU(3)$_C \times$ SU(2)$_L \times$ U(1)$_Y$, 
corresponding to the gauge charge B--L \cite{MM}, where the anomaly 
cancellation requirement implies the existence of three right-handed singlet 
neutrinos.  However it cannot explain the large mixing between the $\nu_\mu$ 
and $\nu_\tau$ states and their small mixing with $\nu_e$, since it treats the 
three flavours on equal footing.  Furthermore, since $m_3 = 0$ is allowed, 
we need only \underline {two} heavy right-handed singlet neutrinos.  Such a 
see-saw model was recently considered by us \cite{MRS}, which was based on 
the gauge group U(1)$_{B-3Le}$, thereby distinguishing the $e$ flavour from 
$\mu$ and $\tau$ in the choice of the gauge group.  We present here a 
\underline {more economical} and \underline {better motivated} model based 
on a slightly different U(1)$_{Y^\prime}$ extension of the SM with 
\be                    
Y^\prime \; = \; B - \textstyle{\frac{3}{2}} \: (L_\mu + L_\tau).             
\label{eq:a6}                    
\ee                    
This U(1) $_{Y^\prime}$ can only be gauged together with the SM if there are 
\underline {two} right-handed singlet neutrinos carrying this charge, as we 
see below.  We now have a reason why $\nu_\mu$ and $\nu_\tau$ are different 
from $\nu_e$, and also why the $\nu_e$ mass is zero.  Contrast this with the 
usual B--L 
model \cite{MM} where there must be three singlets and the B--3L$_e$ model 
\cite{MRS} where there is only one.  In the latter, an extra singlet neutrino 
has to be added by hand, and it must not have any gauge interactions, hence 
its existence is not very well motivated.

The two extra right-handed singlet neutrinos have normal Dirac couplings to 
$\nu_\mu$ and $\nu_\tau$ but suppressed ones to $\nu_e$ because they do so 
through a different Higgs doublet which has a naturally small vacuum 
expectation value (vev) as we see below.  This ensures the desired mixing 
pattern of (\ref{eq:a1}) and (\ref{eq:a3}).  Moreover, one can get the 
induced neutrino masses in the desired range of (\ref{eq:a5}), assuming 
a U(1) symmetry breaking scale of $\sim 10^{12-16}$~GeV.  Thus the model can 
naturally account for the large (small) mixing solutions to the atmospheric 
(solar) neutrino oscillations.                    
                    
The SU(3)$_C \times$ SU(2) $\times$ U(1)$_Y \times$ U(1)$_{Y^\prime}$ gauge 
charges of the quarks and leptons, including the two singlet neutrinos, are 
listed below 
\bea                    
\label{eq:a7}                    
\left (\begin{array}{c} u_i \\ d_i \end{array} \right )_L & \sim & \left (
3, 2, \textstyle{\frac{1}{6}, \frac{1}{3}} \right ); \quad u_{iR} \; \sim 
\; \left (3, 1, \textstyle{\frac{2}{3}}, \frac{1}{3} \right  ); \quad d_{iR} 
\; \sim \; \left (3, 1, \textstyle{\frac{-1}{3}, \frac{1}{3}} \right ); 
\nonumber \\  & & \nonumber \\  \left ( \begin{array}{c} \nu_e \\ e 
\end{array} \right )_L & \sim & \left (1, 2, \textstyle{\frac{-1}{2}}, 0 
\right ); \quad e_R \; \sim \; (1, 1, -1, 0); \nonumber \\  & & \nonumber \\ 
\left ( \begin{array}{c} \nu_\mu \\ \mu \end{array} \right )_L, \left ( 
\begin{array}{c}  \nu_\tau \\ \tau \end{array} \right )_L & \sim & \left (
1, 2, \textstyle{\frac{-1}{2}, \frac{-3}{2}} \right ); \quad \mu_R, \tau_R \; 
\sim \; \left (1, 1, -1, \textstyle{\frac{-3}{2}} \right ); \nonumber \\   
& & \nonumber \\ \nu_{1R}, \nu_{2R} & \sim & \left (1, 1, 0, 
\textstyle{\frac{-3}{2}} \right ).                  
\eea     

The cancellation of anomalies has been discussed in \cite{MR} in the context 
of an analogous U(1) extension of the SM.  Since the number of SU(2)$_L$ 
doublets remain unchanged (even), the global SU(2) chiral gauge anomaly 
\cite{EW} is absent.  The presence of the two right-handed singlet 
neutrinos ensures that the quarks and leptons transform vectorially under 
the U(1)$_{Y^\prime}$.  Consequently the mixed gravitational-gauge anomaly 
\cite{DS} is absent.  It also ensures the absence of the                
[SU(3)$_C$]$^2$ U(1)$_{Y^\prime}$ and [U(1)$_{Y^\prime}$]$^3$ axial-vector 
anomalies \cite{SLA}.  The other axial vector triangle anomalies are 
cancelled as follows                
\bea               
\label{eq:a8}               
[{\rm SU}(2)]^2 {\rm U(1)}_{Y^\prime} & : & (3) (3) (\textstyle{\frac{1}{3}}) 
\: + \:  (2)(\textstyle{\frac{-3}{2}}) \; = \; 0, \\ 
\label{eq:a9} [{\rm U(1)}_{Y^\prime}]^2 {\rm U(1)}_Y & : & (3) (3) 
(\textstyle{\frac{1}{3}})^2 [2 (\textstyle{\frac{1}{6}}) - 
(\textstyle{\frac{2}{3}}) - (\textstyle{\frac{-1}{3}})]  \nonumber \\ 
& & + \: (2) (\textstyle{\frac{-3}{2}})^2 [2 (\textstyle{\frac{-1}{2}}) - 
(-1)] \; = \; 0, \\   \label{eq:a10} {\rm U(1)}_{Y^\prime} 
[{\rm U(1)}_Y]^2 & : & (3) (3) (\textstyle{\frac{1}{3}}) [2 
(\textstyle{\frac{1}{6}})^2 - (\textstyle{\frac{2}{3}})^2 - 
(\textstyle{\frac{-1}{3}})^2 ] \nonumber \\  & & + \: (2) 
(\textstyle{\frac{-3}{2}}) [2 (\textstyle{\frac{-1}{2}})^2 - (-1)^2] \; = \; 
0,               
\eea               
where the first two entries in each equation refer to numbers of quark 
colours and generations.  Thus the $Y^\prime$ symmetry can be gauged 
along with the others.           
           
The minimal scalar sector of the model consists of the SM Higgs doublet and 
a neutral singlet,           
\be           
\left ( \begin{array}{c} \phi^+ \\ \phi^0 \end{array} \right ) \; \sim \; 
\left (1, 2, \textstyle{\frac{1}{2}}, 0 \right ), \quad \chi^0 \; \sim \; 
(1, 1, 0, 3). \label{eq:a11}           
\ee           
The latter couples to the singlet pairs $\nu_i \nu_j$, while the former is 
responsible for their Dirac couplings to $\nu_\mu$ and $\nu_\tau$.  This 
will be adequate for atmospheric neutrino oscillations but not for solar 
neutrino, as $\nu_e$ will completely decouple from the other neutrinos.  
Therefore we shall assume another Higgs doublet and a singlet,           
\be           
\left ( \begin{array}{c} \eta^+ \\ \eta^0 \end{array} \right ) \; \sim \; 
\left (1, 2, \textstyle{\frac{1}{2}, \frac{-3}{2}} \right ), \quad \zeta^0 \; 
\sim \; \left (1, 1, 0, \textstyle{\frac{-3}{2}} \right ).  \label{eq:a12} 
\ee           
The doublet shall account for the suppressed Dirac couplings of the singlet 
neutrinos to $\nu_e$.  The singlet does not couple to the fermions; but is 
required to avoid an unwanted pseudo-Goldstone boson \cite{MR}.  This comes 
about because there are 3 global U(1) symmetries, corresponding to rotating 
the phases of $\phi, \eta$ and $\chi^0$ independently in the Higgs potential, 
while only 2 local U(1) symmetries get broken.  The addition of the singlet 
$\zeta^0$ introduces two more terms in the Higgs potential, $\eta^\dagger 
\phi \zeta^0$ and $\chi^0 \zeta^0 \zeta^0$, so that the extra global symmetry 
is eliminated.           
           
Both $\chi^0$ and $\zeta^0$ are expected to acquire large vev's and masses 
at the scale of the U(1)$_{Y^\prime}$ symmetry breaking.  In contrast, the 
doublet $\eta$ is required to have a positive mass squared term in order to 
avoid SU(2) breaking at this scale.  Nonetheless it can acquire a small but 
non-zero vev as the SU(2) symmetry gets broken \cite{EM}.  This can be 
estimated from the relevant part of the Higgs potential           
\be           
m_\eta^2 \eta^\dagger \eta \: + \: \lambda (\eta^\dagger \eta) (\chi^\dagger 
\chi) \: + \:  \lambda^\prime (\eta^\dagger \eta) (\zeta^\dagger \zeta) \: 
- \: \mu \eta^\dagger \phi \zeta. \label{eq:a13}           
\ee           
Although we start with a positive mass squared term for $\eta$, after 
minimisation of the potential we find that this field has acquired a small 
vev,          
\be          
\langle \eta \rangle \; = \; \mu \langle \phi \rangle \: \langle \zeta \rangle / M_\eta^2,          
\label{eq:a14}          
\ee          
where $M_\eta^2 = m_\eta^2 + \lambda \langle \chi \rangle^2 + \lambda^\prime 
\langle \zeta \rangle^2$ represents the physical mass of $\eta$ and $\langle 
\phi \rangle \simeq 10^2$~GeV.  The size of the soft term is bounded by the 
$Y^\prime$  symmetry breaking scale, i.e. $\mu \leq \langle \zeta \rangle$.  
In order to account for the small mixing angle of $\nu_e$ (\ref{eq:a3}), 
we shall require          
\be          
\langle \eta \rangle/\langle \phi \rangle \; \sim \; 10^{-2}.          
\label{eq:a15}          
\ee          
This would correspond to assuming $\mu \sim \langle \zeta \rangle/100$, or 
alternatively $\mu \sim \langle \zeta \rangle$ and $M_\eta \simeq m_\eta 
\simeq 10 \langle \zeta \rangle$.  In either case one can get the desired 
vev with a reasonable choice of the mass parameters.          
          
As usual we shall be working in the basis where the charged lepton 
mass matrix, arising from their couplings to the SM Higgs boson $\phi$, 
is diagonal.  This defines the flavour basis for the doublet neutrinos.  
Since the two singlet neutrinos are decoupled from the charged leptons, 
their Majorana mass matrix can be diagonalised independently.  We shall 
denote their mass eigenvalues as $M_1$ and $M_2$.  While the overall size 
of these masses will be at the $Y^\prime$ symmetry breaking scale,           
we shall assume a modest hierarchy between them,          
\be          
M_1/M_2 \; \sim \; 1/20,          
\label{eq:a16}          
\ee          
in order to account for the desired mass ratio for the doublet neutrinos 
(\ref{eq:a5}).  The above hierarchy between the singlet neutrino masses 
compares favourably with those observed in the quark and charged lepton 
sectors.          
          
Thus we have the following $5 \times 5$ neutrino mass matrix in the basis 
$(\nu_e, \nu_\mu, \nu_\tau, \nu_1^c, \nu_2^c)$:          
\be          
M \; = \; \left ( \begin{array}{ccccc}          
0 & 0 & 0 & f_e^1 \langle \eta \rangle & f_e^2 \langle \eta \rangle \\ 
& & & & \\          
0 & 0 & 0 & f_\mu^1 \langle \phi \rangle & f_\mu^2 \langle \phi \rangle \\  
& & & & \\          
0 & 0 & 0 & f_\tau^1 \langle \phi \rangle & f_\tau^2 \langle \phi \rangle \\  
& & & & \\          
f_e^1 \langle \eta \rangle & f_\mu^1 \langle \phi \rangle & f_\tau^1 \langle 
\phi \rangle & M_1 & 0 \\         
& & & & \\          
f_e^2 \langle \eta \rangle & f_\mu^2 \langle \phi \rangle & f_\tau^2 \langle 
\phi \rangle & 0 & M_2 \\ \end{array} \right ), \label{eq:a17}          
\ee        
where $\nu_{1,2}^c$ denote antiparticles of the right-handed singlet 
neutrinos and the $f$'s are the Higgs Yukawa couplings.  The induced 
mass matrix for the doublet neutrinos is easy to calculate in our basis of 
a diagonal Majorana mass matrix.  It is given by the see-saw formula in 
this basis,        
\be        
m_{ij} \; = \; \frac{D_{1i} D_{1j}}{M_1} \: + \: \frac{D_{2i} D_{2j}}{M_2}, 
\label{eq:a18}        
\ee        
where $i, j$ denote the 3 neutrino flavours and $D$ represents the 
$2 \times 3$ Dirac mass matrix at the bottom left of (\ref{eq:a17}).  
We get        
\be        
m \; = \; \left ( \begin{array}{ccc}        
c_1^2 + c_2^2 & c_1 a_1 + c_2 a_2 & c_1 b_1 + c_2 b_2 \\        
& & \\        
c_1 a_1 + c_2 a_2 & a_1^2 + a_2^2 & a_1 b_1 + a_2 b_2 \\        
& & \\        
c_1 b_1+ c_2 b_2 & a_1 b_1 + a_2 b_2 & b_1^2 + b_2^2 \end{array} \right ), 
\label{eq:a19}        
\ee        
where        
\be       
a_{1,2} \; = \; \frac{f_\mu^{1,2} \langle \phi \rangle}{\sqrt{M_{1,2}}}, 
\quad        
b_{1,2} \; = \; \frac{f_\tau^{1,2} \langle \phi \rangle}{\sqrt{M_{1,2}}}, 
\quad  
c_{1,2} \; = \; \frac{f_e^{1,2} \langle \eta \rangle}{\sqrt{M_{1,2}}}.       
\label{eq:a20}       
\ee        
       
We shall assume all the Yukawa couplings to be of the same order of 
magnitude, which means that the elements of a mass matrix arising from the 
same Higgs vev are expected to be of similar size.  There is of course no 
conflict between the assumption of democratic mass matrix elements and 
hierarchical mass eigenvalues \cite{FX}.  In fact the latter requires 
large cancellations in the determinant, which in turn implies        
democratic elements of the mass matrix.  This appears to be a reasonable 
assumption, although we shall use it only for a limited purpose -- i.e. to 
ensure that the hierarchies resulting from the ratios of the Higgs vev's 
(\ref{eq:a15}) and the singlet mass eigenvalues (\ref{eq:a16}) are not 
washed out by violent fluctuations in the Higgs couplings.  Then these 
hierarchies imply       
\be       
a_1, b_1 \; \gg \; a_2, b_2, c_1 \; \gg \; c_2.       
\label{eq:a21}       
\ee       
This leads to a texture of the mass matrix (\ref{eq:a19}), where the $\{11\}$
element is doubly suppressed and the remaining elements of the first row 
and first column are singly suppressed \cite{AF}.  It is a reflection of 
the hierarchy (\ref{eq:a15}) in the Dirac mass matrix, which will show up 
in the hierarchy of the two mixing angles (\ref{eq:a1}) and (\ref{eq:a3}).  
On the other hand the hierarchy (\ref{eq:a16}) of Majorana mass eigenvalues 
will be reflected in a similar hierarchy between the non-zero eigenvalues of 
(\ref{eq:a19}), which correspond to the two neutrino masses of (\ref{eq:a5}). 
       
One can easily check that the determinant of the mass-matrix (\ref{eq:a19}) 
vanishes, so that one of its eigenvalues is zero.  The other two eigenvalues 
are       
\bea    
\label{eq:a22}  
&& m_{1,2} = \frac{1}{2} \biggl [a_1^2 + a_2^2 + b_1^2 + b_2^2 + c_1^2 + 
c_2^2 \biggr .  \\  && \pm \; \biggl . \sqrt{(a_1^2 + a_2^2 
+ b_1^2 + b_2^2 + c_1^2 + c_2^2) ^2 - 4 \left \{ (a_1 b_2 - b_1 a_2)^2 
+ (a_1 c_2 - c_1 a_2)^2 + (b_1 c_2 - c_1 b_2)^2 \right \}} \biggr ]. 
\nonumber       
\eea     
>From (\ref{eq:a21}) and (\ref{eq:a22}) we get   
\bea   
\label{eq:a23}   
m_1 & \simeq & a_1^2 \: + \: b_1^2, \\   
& & \nonumber \\   
\label{eq:a24}   
m_2 & \simeq & \frac{(a_1 b_2 \: - \: a_2 b_1)^2}{a_1^2 \: + \: b_1^2},    
\eea   
i.e.   
\be   
m_2/m_1 \; \sim \; M_1/M_2.   
\label{eq:a25}   
\ee   
Thus the assumed hierarchy of the Majorana masses (\ref{eq:a16}) do account 
for the relative size of the two neutrino masses of (\ref{eq:a5}).  Moreover 
the required size of $m_1$ or $m_2$ will give the overall scale of the 
$Y^\prime$ symmetry breaking Majorana mass, i.e.   
\be   
M_2 \; \sim \; (f_{\mu, \tau}^2)^2 \: \langle \phi \rangle^2/m_2 \; \sim \; (f_{\mu, \tau}^2)^2 \: 10^{16}~{\rm GeV}.   
\label{eq:a26}   
\ee   
Assuming the size of the Yukawa couplings to be similar to the top Yukawa    
coupling ($\sim$ 1), we then have   
\be   
M_2 \; \sim \; 10^{16}~{\rm GeV},   
\label{eq:a27}   
\ee   
i.e. close to a possible grand unification scale.  On the other hand, 
assuming the Yukawa couplings to be similar in size to that of thr $\tau$ 
lepton $(\sim 10^{-2})$ would imply   
\be   
M_2 \; \sim \; 10^{12}~{\rm GeV}.   
\label{eq:a28}   
\ee   
Thus within the lattitude of the Yukawa coupling given above, the 
$Y^\prime$ symmetry breaking scale could be anywhere in the range 
$10^{12-16}$~GeV.   
   
Finally we can calculate the eigenvectors corresponding to the three 
eigenvalues, $m_1, m_2$ and $m_3 (= 0)$.  This gives the following mixing 
matrix connecting the flavour eigenstates to the mass eigenstates, written 
in increasing order of mass :   
\be   
\left ( \begin{array}{c} \nu_e \\ \nu_\mu \\ \nu_\tau \end{array} \right ) \; 
= \; \left (  \begin{array}{ccc} 1 & \displaystyle{\frac{{-c_2}{\sqrt{a_1^2 
+ b_1^2}}}{a_1 b_2 - b_1 a_2}} &  
\displaystyle{\frac{c_1}{\sqrt{a_1^2 + b_1^2}}} \\   & & \\   
\displaystyle{\frac{b_1 c_2 - c_1 b_2}{a_1 b_2 - b_1 a_2}} &   
\displaystyle{\frac{b_1}{\sqrt{a_1^2 + b_1^2}}} &    
\displaystyle{\frac{a_1}{\sqrt{a_1^2 + b_1^2}}} \\  & & \\   
\displaystyle{\frac{c_1 a_2 - a_1 c_2}{ a_1 b_2 - b_1 a_2}} &   
\displaystyle{\frac{-a_1}{\sqrt{a_1^2 + b_1^2}}} &    
\displaystyle{\frac{b_1}{\sqrt{a_1^2 + b_1^2}}} \end{array} \right ) \; 
\left (  \begin{array}{c} \nu_3 \\ \nu_2 \\ \nu_1 \end{array} \right ).   
\label{eq:a29}   
\ee   
  
The large mixing angle, responsible for atmospheric neutrino oscillations, 
corresponds to  
\be  
\tan \theta_{\mu \tau} \; = \; a_1/b_1 \; = \; f_\mu^1/f_\tau^1,  
\label{eq:a30}  
\ee  
i.e. it is given by the ratio of the SM Higgs Yukawa couplings to $\nu_\mu$ 
and   $\nu_\tau$ along with the lighter singlet.  Assuming these Yukawa 
couplings to be equal implies maximal mixing, $\theta_{\mu \tau} = 45^\circ$. 
Moreover, any value of their ratio in the range  
\be  
0.6 \; < \; f_1/f_2 \; < \; 1.6  
\label{eq:a31}  
\ee  
will ensure the large mixing angle (\ref{eq:a1}) required by data, which 
corresponds to $32^\circ < \theta_{\mu \tau} < 58^\circ$.  Thus one can get 
the required mixing angle for atmospheric neutrino oscillation without any 
fine tuning of the Yukawa couplings.  
  
The small mixing angle, responsible for solar neutrino oscillations, 
corresponds to the mixing of the $\nu_e$ with the lighter mass eigenstate 
$\nu_2$, i.e.  
\be  
\sin \theta_{e - \mu/\tau} \; \simeq \; \frac{{c_2}{\sqrt{a_1^2 + b_1^2}}}
{a_1 b_2 - b_1 a_2} \; \sim \;   \frac{\langle \eta \rangle}{\langle \phi 
\rangle}.  \label{eq:a32}  
\ee  
Thus the ratio (\ref{eq:a15}) of the two Higgs vev's can account for the 
required size of the mixing angle (\ref{eq:a3}), i.e.  
\be  
\sin \theta_{e - \mu/\tau} \; = \; (1.6 - 5) \: \times \: 10^{-2}.  
\label{eq:a33}  
\ee  
It should be noted that in this model, one also expects a similar size of 
$\nu_e$  mixing with the heavier mass eigenstate $\nu_1$.  This is allowed 
by all current experiments, including CHOOZ \cite{CHOOZ}, although it has 
been assumed to be zero in some mixing models.  Hopefully this mixing angle 
can be probed by future reactor and long baseline accelerator experiments.  
  
Notice that $\eta$ also couples $e_R$ to $\mu_L$ and $\tau_L$, 
which introduces small non-diagonal elements in the charged lepton mass 
matrix.  However, as shown in \cite{MRS}, its contribution to the $\nu_e$ 
mixing angle is very small $(\sin \theta_{e - \mu/\tau} \leq 10^{-3})$.  
The theoretical origin of our proposed U(1)$_{Y^\prime}$ is not obvious.  
It spans all three quark families but only two lepton families.  A 
possibility is that at the putative grand unification scale, what exists is 
a remnant of a string theory which already breaks down to the SM together 
with this extra U(1). The low-energy consequence of our model is identical 
to that of the SM, including the effective Higgs sector, except for neutrino 
masses.
  
In summary, we have considered a see-saw model based on a new U(1) extension 
of the SM gauge group, corresponding to the gauge charge $B - 3/2 (L_\mu + 
L_\tau)$.  The requirement of anomaly cancellation implies the existence of 
two right-handed singlet neutrinos, carrying this gauge charge, which have 
normal Dirac couplings to $\nu_\mu$ and $\nu_\tau$, but suppressed ones to 
$\nu_e$.  Consequently they induce see-saw masses to two doublet neutrino 
states, which are large admixtures of $\nu_\mu$ and $\nu_\tau$ with small 
$\nu_e$ components.  Moreover, one can get the right size of these neutrino 
masses for explaining the large (small) mixing solutions to the atmospheric 
(solar) neutrino oscillations, if the scale of this U(1) symmetry breaking 
is in the range of $10^{12-16}$~GeV.  The necessity of two and only two 
singlet neutrinos of the $\mu$ and $\tau$ variety in this model tells us 
why $\nu_\mu - \nu_\tau$ mixing is large and why $\nu_e$ is massless. 
Thus it represents what appears to be a minimal see-saw model for explaining 
these oscillations. 
\\  
  
\noindent {\large \bf Acknowledgements}  
  
DPR acknowledges the hospitality of Grey College and the Physics Department 
of the University of Durham and financial support from PPARC (UK). 
EM acknowledges the hospitality of CERN where this work was completed.  
Support in part was provided by the U.~S.~Department of Energy under 
Grant No.~DE-FG03-94ER40837.


\newpage                       
\begin{thebibliography}{xx}                       
\bibitem{SKC1} Super-Kamiokande Collaboration: Y.~Fukuda et al., Phys.~Rev.~Lett
.~{\bf 81} (1998) 1562; Phys.~Lett.~{\bf B433} (1998) 9 and ~{\bf B436} (1998) 33; \\ 
T.~Kajita, Talk presented at Neutrino-98, Takayama, Japan (1998).
%
\bibitem{SKC2} Super-Kamiokande Collaboration:  Y.~Fukuda et al., 
Phys.~Rev.~Lett.~{\bf 81} (1998) 1158; \\  
Talk by Y.~Suzuki at Neutrino-98, Takayama, Japan (1998).  
%
\bibitem{LW} L.~Wolfenstein, Phys.~Rev.~{\bf D17} (1978) 2369; \\        
S.P.~Mikheyev and A.Yu.~Smirnov, Sov.~J.~Nucl.~Phys.~{\bf 42} (1986) 913.     
%
\bibitem{BKS} J.N.~Bahcall, P.J.~Krastev and A.Yu.~Smirnov, Phys.~Rev.~{\bf D58
} (1998) 096016; \\                       
N.~Hata and P.G.~Langacker, Phys.~Rev.~{\bf D56} (1997) 6107.           
%
\bibitem{BA} B.~Allanach, {\tt hep-ph/9806294}; \\                       
V.~Barger, S.~Pakvasa, T.J.~Weiler and K.~Whisnant, Phys.~Lett.~{\bf B437} (1998) 107; \\  
V.~Barger, T.J.~Weiler and K.~Whisnant, {\tt hep-ph/9807319}; \\              
E.~Ma, {\tt hep-ph/9807386} (Phys. Lett. {\bf B}, in press); \\                       
Y.~Nomura and T.~Yanagida, {\tt hep-ph/9807325}; \\                       
A.~Joshipura, {\tt hep-ph/9808261}; \\                       
J.~Ellis et al., {\tt hep-ph/9808301}; \\                       
E.J.~Chun et al., {\tt hep-ph/9807327}; \\                       
A.~Joshipura and S.~Vempati, {\tt hep-ph/9808232}; \\                       
U.~Sarkar, {\tt hep-ph/9808277}; \\                       
G.~Cleaver, M.~Cvetic, J.R.~Espinosa, L.~Everett and P.~Langacker,         
Phys.~Rev.~{\bf D57} (1998) 2701; \\                       
H.~Georgi and S.L.~Glashow, {\tt hep-ph/9808293}; \\                       
R.N.~Mohapatra and S.~Nusinov, {\tt hep-ph/9808301}; \\                       
R.~Barbieri, L.J. Hall and A.~Strumia, {\tt hep-ph/9808333}; \\         
R.~Barbieri et al., {\tt hep-ph/9807235}; \\                       
Y.~Grossman, Y.~Nir and Y.~Shadmi, {\tt hep-ph/9808355}.                 
%
\bibitem{FX} H.~Fritzsch and Z.~Xing, {\tt hep-ph/9808272}.            
%
\bibitem{AF} G.~Altarelli and F.~Feruglio, {\tt hep-ph/9807353} and {\tt  
hep-ph/9809596}; \\     
J.~Sato and T.~Yanagida, Phys.~Lett.~{\bf B430} (1998) 127 and      
{\tt hep-ph/9809307}; \\     
J.~Elwood, N.~Irges and P.~Raymond, {\tt hep-ph/9807226}.      
%
\bibitem{EM} E.~Ma and U.~Sarkar, Phys.~Rev.~Lett.~{\bf 80} (1998) 5716;\\
E. Ma, Phys.~Rev.~Lett.~{\bf 81} (1998) 1171..  
%
\bibitem{DPTV} M.~Drees, S.~Pakvasa, X.~Tata and T.~ter~Veldhuis,     
Phys.~Rev.~{\bf D57} (1998) 5335; \\                       
B.~Mukhopadhyaya, S.~Roy and F.~Vissani, {\tt hep-ph/9808265}.   
%
\bibitem{SFK} S.F.~King, {\tt hep-ph/9806440}; \\                  
S.~Davidson and S.F.~King, {\tt hep-ph/9808296}.                       
%
\bibitem{GMRS} M~Gell-Mann, P.~Ramond and R.~Slansky, in {\it Supergravity},  
Proceedings of the Workshop, Stony Brook, New York, 1979, ed.~by P.~van        
Nieuwenhuizen and D.~Freedman (North-Holland, Amsterdam); \\     
T.~Yanagida, in Proceedings of the Workshop on Unified Theories and Baryon     
Number in the Universe, Tsukuba, Japan, edited by A.~Sawada and A.~Sugamoto    
(KEK Report No. 79-18, Tsukuba) 1979.                      
%
\bibitem{MM} R.E.~Marshak and R.N.~Mohapatra, Phys.~Lett.~{\bf B91} (1980)     
222.                      
%
\bibitem{MRS} E.~Ma, D.P.~Roy and U.~Sarkar, {\tt hep-ph/9810309} (Phys. 
Lett. {\bf B}, in press) .   
%
\bibitem{MR} E.~Ma, Phys.~Lett.~{\bf B433} (1998) 74; \\                      
E.~Ma and D.P.~Roy, Phys.~Rev.~{\bf D58} (1998) 095005.                      
%
\bibitem{EW} E.~Witten, Phys.~Lett.~{\bf B117} (1982) 324.           
%
\bibitem{DS} R.~Delbourgo and A.~Salam, Phys.~Lett.~{\bf B40} (1972) 381; \\ 
T.~Eguchi and P.G.O.~Freund, Phys.~Rev.~Lett.~{\bf 37} (1976) 1251; \\       
L.~Avarez-Gaume and E.~Witten, Nucl.~Phys.~{\bf B234} (1984) 269.      
%
\bibitem{SLA} S.L.~Adler, Phys.~Rev.~{\bf 177} (1969) 2426; \\    
J.S.~Bell and R.~Jackiw, Nuovo Cimento {\bf A60} (1969) 47; \\   
W.A.~Bardeen, Phys.~Rev.~{\bf 184} (1969) 1848.                      
%
\bibitem{CHOOZ} CHOOZ Collaboration:  M.~Apollonio et al., Phys.~Lett.~{\bf 
B420} (1998) 397.                      
\end{thebibliography}
\end{document}